%% file: bare_jrnl.tex
\documentclass[times]{stvrauth}
\pdfoutput=1
\usepackage{graphicx}
\usepackage{tabularx}
\usepackage[title]{appendix}
\usepackage{placeins}
\usepackage{amsmath}
\usepackage{todonotes}
\usepackage{verbatim}
\usepackage{etoolbox}
\usepackage{soul}
\usepackage{subcaption}
\usepackage{adjustbox}
\usepackage{caption}
\usepackage{subcaption}
\usepackage{float}
\usepackage{subfloat}
\usepackage{url}
\usepackage{comment}
\usepackage{multirow}
\usepackage{array}
\usepackage{enumitem}
\usepackage{colortbl}
\usepackage{lmodern}
\usepackage[utf8]{inputenc}
\usepackage[T1]{fontenc}
\usepackage{flexisym}
\usepackage{xcolor}

\newcolumntype{P}[1]{>{\centering\arraybackslash}p{#1}}

\newcolumntype{M}[1]{>{\centering\arraybackslash}m{#1}}
\begin{document}

\textbf{\huge \color{red}\centering This Preprint submitted to Journal of Software Testing Verification and Reliability}
\title{Metamorphic Relation Prioritization for Effective Regression Testing}

\author{Madhusudan Srinivasan\affil{1} and Upulee Kanewala\affil{2} \corrauth}

\address{
\affilnum{1} Gianforte School of Computing, Montana State University, Bozeman, MT, USA. \break \affilnum{2} School of Computing, University of North Florida, Jacksonville, FL, USA.}

\corraddr{Upulee Kanewala, University of North Florida, Jacksonville, FL, USA.
\break
E-mail:upulee.kanewala@unf.edu }

\begin{abstract}
 Metamorphic testing (MT) is widely used for testing programs that face the oracle problem. It uses a set of metamorphic relations (MRs), which are relations among multiple inputs and their corresponding outputs to determine whether the program under test is faulty. Typically, MRs vary in their ability to detect faults in the program under test, and some MRs tend to detect the same set of faults. In this paper, we propose approaches to prioritize MRs to improve the efficiency and effectiveness of MT for regression testing. We present two MR prioritization approaches: (1) fault-based and (2) coverage-based. To evaluate these MR prioritization approaches, we conduct experiments on three complex open-source software systems. Our results show that the MR prioritization approaches developed by us significantly outperform the current practice of executing the source and follow-up test cases of the MRs in an ad-hoc manner in terms of fault detection effectiveness. Further, fault-based MR prioritization leads to reducing the number of source and follow-up test cases that needs to be executed as well as reducing the average time taken to detect a fault, which would result in saving time and cost during the testing process.


\end{abstract}


\maketitle

\input{Introduction.tex}

\input{RelatedWork.tex}

\input{MRPrioritization.tex}
\input{Evaluation.tex}
\input{Result.tex}

\input{ThreattoValidity.tex}
\input{Conclusion.tex}


\ack
The authors would like to thank T.Y. Chen for his valuable ideas, discussions and feedback on this work especially on the coverage-based MR prioritization. The authors would also like to thank Prashanta Saha for the execution of IBk program. This  work  is  supported  by  the award  number  1656877 from the National Science Foundation. Any Opinions, findings and conclusions or recommendations expressed in this material are those of  the  author(s)  and  do  not  necessarily  reflect  those of the National Science Foundation.

\bibliographystyle{wileyj}
\bibliography{references}
\appendices
\input{appendix.tex}

\end{document}

%% file: Introduction.tex
\section{Introduction}
\label{section:Introduction}
A \emph{test oracle} is an essential component of testing that is used to distinguish between correct and incorrect behavior of the system under test (SUT). However, for complex systems such as scientific software or machine learning applications developing test oracles is practically difficult~\cite{Kanewala20141219}. This is known as the \emph{oracle problem} in software testing~\cite{barr2015oracle}. Metamorphic testing (MT) is an approach that is widely used to alleviate the oracle problem. Instead of verifying the output produced by a test case using a test oracle, MT involves verifying the outputs of multiple executions of the SUT and their corresponding inputs satisfy a given~\emph{metamorphic relation (MR)}~\cite{chen2018metamorphic,segura2018metamorphic}. Formally, a MR is a necessary property of the SUT and is specified over multiple inputs and their corresponding outputs~\cite{chen2018metamorphic}.  The typical process for conducting MT is as follows: 
 \begin{enumerate}
\item Identify MRs for the SUT using its specification or the programmer's knowledge about the SUT.
\item Create \emph{source test cases} and execute them on the SUT. Existing test cases can be used as source test cases or can be generated using test generation approaches such as random generation.  

\item Construct the \emph{follow-up test cases} based on the MRs and execute them on the SUT.

\item If the source and follow-up test inputs and outputs do not satisfy a given MR, then the SUT is faulty.

\end{enumerate}
MT has led to the detection of previously unknown faults in many applications across various domains. Such examples include the detection of 147 confirmed bugs in two-open source C compilers, GCC and LLVM using the equivalence preservation MR~\cite{tao2010automatic}. More recently, MT was used to reveal hidden defects in two widely used citation database systems, Scopus and Web of Science particularly when handling hyphenated paper titles~\cite{zhou2019metamorphic}. Further, Zhou et al~\cite{zhou2018metamorphic} proposed symmetry metamorphic relation pattern to derive multiple concrete MRs that lead to revealing previously unknown failures in widely used applications; in particular this lead to revealing a fault in Google map navigation when the origin and destination were swapped. Additionally, MT was applied to conduct a large scale empirical study using four major web search engines: Google, Bing, Chinese Bing, and Baidu~\cite{zhou2015metamorphic}. The experimental results of this study revealed that MT could reveal various types of failures that impact the quality of the results generated by a search~\cite{zhou2015metamorphic}.  

With the successful applications of MT in different domains, it becomes important to develop techniques that allow using MT effectively for \emph{regression testing}. Regression testing is conducted every time a program is modified in order to make sure that no new bugs were introduced during the modification process~\cite{ammann2016introduction}. With the wide adaptation of agile software development methods, regression testing has to be done more frequently with releases of successive versions. Therefore, in such a development environment, it becomes beneficial to use  efficient and effective \emph{MR prioritization} approaches. In addition, previous studies have shown that typically five to nine MRs are used for testing~\cite{segura2016survey} and the cost (in terms of time and resources) of the MT process is proportional to the number of MRs used for testing. SUT can have MRs with multiple source and follow-up test cases. Thus, as the number of MRs increases, the number of test cases can increase exponentially and as a result, time taken to complete execution also increases. Further, machine learning and bioinformatics programs can consume a significant amount of time and resources to execute when used with typical data. Therefore, it becomes beneficial to prioritize MRs even with a smaller MR set size. Also often, individual MRs significantly differ in their fault detection effectiveness and some MRs tend to detect the same type of faults~\cite{srinivasan2018quality}. Therefore, selecting a diverse set of MRs enhances the fault detection effectiveness of MT~\cite{liu2014effectively,chen2004case,cao2013correlation} and would help to save resources during regression testing. However, while Cao et al.~\cite{cao2013correlation} work provides
hints/quantitative guidelines towards this direction, there are no existing automated methods for selecting or prioritizing MRs based on their potential fault detection effectiveness.  

To this end, in this work, we propose methods for \emph{MR prioritization} to improve the efficiency and effectiveness of MT for regression testing. The MR prioritization methods proposed in this work utilize the fault detection information and the coverage information collected while conducting MT on the previous versions of the SUT. Thus, these methods do not require any additional executions of the SUT. Then, a light weight computation is used to derive the prioritized MR ordering using the collected information. Therefore, the cost of applying the MR prioritization is very low. However, our results show that by utilizing MR prioritization the number of MRs that needs to be executed on the current version of SUT can be reduced significantly and the time taken to detect a fault could be reduced by up to 68\%.



We make the following contributions in this work:
\begin{enumerate}
    \item Propose two novel automated methods for \emph{MR prioritization}. The first method is \emph{fault-based MR prioritization}, where fault detection effectiveness of MRs are used to prioritize them. The second method is the \emph{coverage-based MR prioritization}, where the statement and branch coverage information of the MRs are used to prioritize them.
    \item  Evaluate the  effectiveness of our proposed MR prioritization methods using three complex real-world systems from diverse domains and perform a comparison against two baseline approaches: (1) Random baseline: represents the current practice of executing the MRs in a random order, and (2) Optimal ordering: represents the best possible ordering of the MRs for detecting a given set of faults.
            \end{enumerate}
            

We also make the following conclusions based on our experimental results:
\begin{enumerate}
    \item Developed MR prioritization approaches report significant improvements in fault detection up to 266.66\% compared to the current practice of random execution of MRs. Also, using fault-based MR prioritization, the effective number of MRs that need to be executed could be reduced from 8 to 4 for some programs, resulting in significant savings in testing resources. Further, using the fault-based MR prioritization, the average time taken to detect faults could be reduced by up to 68\%. Here the fault detection time refers to the time taken to execute the source and follow-up test cases of the MRs until the fault is revealed.
    \item Among the developed MR prioritization approaches, fault-based prioritization is the most effective approach in terms of fault detection. It performs closer to the optimal ordering of MRs.
\end{enumerate}

Therefore we conclude that our MR prioritization approaches can be used to improve the effectiveness of MT over the current practice of random execution of MRs in a regression testing environment. In particular, utilizing MR prioritization would lead to reducing the number of MRs that needs to be executed and reducing the time taken to detect faults during regression testing. 


The remainder of the paper is organized as follows: In Section~\ref{section:related work}, work related to MR prioritization is presented. In Section~\ref{section:MR Prioritization}, the proposed approaches for MR prioritization is discussed in detail. In Section~\ref{section:Eval}, the validation procedure for the empirical study is discussed.  In Section~\ref{section:Results&Discussion}, results and answers to the research questions are presented. In Section~\ref{section:ThreattoValidity}, potential threats to validity are discussed. In Section~\ref{Section:Conclusion}, a discussion of conclusions is presented.

%% file: RelatedWork.tex
\section{Related Work}
\label{section:related work}
To the best of our knowledge, this is the first work on developing automated methods for prioritizing MRs in the context of regression testing. However, there are some previous works on prioritizing test cases for regression testing, selecting effective MRs and selecting expected value oracles. We summarize this work here and explain how they differ from our work.
In previous work, mutants and coverage information have been used for test case prioritization. Gregg et al.~\cite{rothermel2001prioritizing} proposed a technique for using test execution information to prioritize test cases for regression testing. This technique orders the test cases based on total coverage of code and also based on coverage of code previously not covered. The test cases are also ordered based on the ability to reveal faults in the code. The experiment result suggests that these techniques can improve the rate of fault detection. Luo et al.~\cite{lou2015mutation} proposed a mutation-based test case prioritization method for software evolution. In this approach, mutants are generated on the statements that are changed between the previous version and the current version of the program. Then, the test cases are prioritized based on the ability of the test cases to detect these mutants. Shin et al.~\cite{shin2019empirical} proposed a mutation-based test case prioritization approach that utilizes a diversity of mutants. They compared the traditional kill-only method against the diversity aware mutation technique under greedy, hybrid, and multi-objective prioritization schemes. Their empirical results show that there is no single superior test case prioritization technique, and combining the kill-only technique and the diversity aware technique improves the effectiveness of test case prioritization. Nardo et al.~\cite{di2015coverage} conducted a case study to evaluate four common coverage-based test case prioritization approaches on a real-world industrial system. Their results show that the prioritization technique using additional coverage outperformed random ordering. These test case prioritization approaches involve the execution of individual test cases in a test suite and prioritizing them based on various techniques and criteria. In comparison, here we focus on prioritizing MRs, which are used for testing programs that face the oracle problem, as opposed to prioritizing test cases. 

Chen et al.~\cite{chen2004case} proposed guidelines for the selection of good MRs. In order to develop guidelines for MR selection, a case study was conducted on the shortest path program that implements the Dijkstra's algorithm. The experiment results suggest that the knowledge of the domain is not adequate for distinguishing good MRs, rather a good MRs make multiple executions of the SUT that are different from each other. Cao et al.~\cite{cao2013correlation} proposed six metrics to measure the dissimilarity between source and follow-up test cases. They conducted an empirical study to measure the dissimilarity between source and follow-up test case executions using these measures. This approach does not employ the prioritization of MRs; however, it helps in determining the effectiveness of the MRs. Mayer and  Guderlei~\cite{mayer2006empirical} conducted an empirical study to evaluate the usefulness of MRs. Based on the experiment result, the authors devise several criteria to asses the quality of the MRs based on the potential usefulness. These criteria were defined based on the different execution paths undertaken by the source and follow-up test cases. While this work provides some directions into selecting useful MRs, they do not directly address MR prioritization. In particular, we focus on developing automated methods for MR prioritization.

Gay et al.~\cite{gay2015automated} proposed a mutation-based approach for selecting a set of internal and external variables to be monitored during testing. In this approach, the variables are ranked based on the mutation killing rate. However, this approach would require the tester to specify the expected values for the variables selected as the oracles. Specifying the expected value for variables may not be possible for programs that face the oracle problem. In this work we focus on prioritizing MRs.
 

%% file: MRPrioritization.tex
\section{MR Prioritization}
\label{section:MR Prioritization}
In this work we propose two approaches for prioritizing MRs: (1) fault-based, and (2) coverage-based. Below we describe these approaches in detail.

\subsection{Fault-based MR Prioritization}
\label{Mutation-based approach}


Let the current version of the SUT be $v_k$. Fault-based MR prioritization utilizes fault detection information of MRs collected while testing the previous version $v_{k-1}$ of the SUT to prioritize MRs for testing $v_k$. 
\begin{enumerate}
    \item Let the set of source test cases used for testing the version $v_{k-1}$ of the SUT  be the \emph{prioritizing source test cases} $(T_{sp})$.
    \item Let the set of follow-up test cases used for testing the version $v_{k-1}$ of the SUT be the \emph{prioritizing follow-up test cases} $(T_{fp})$. 
    \item Let the set of faults detected in version $v_{k-1}$ be the \emph{prioritizing set of faults} $F_p$. For each fault $f \in F_p$, log whether each $MR_i$ revealed $f$ when executed with $(T_{sp})$ and $(T_{fp})$. 
    \item Use the following greedy approach to create the prioritized ordering of the MRs:
        \begin{enumerate}
        \item \label{one} Select the MR that revealed the highest number of faults in $F_p$ and place it in the prioritized MR ordering. If there are multiple MRs with the same highest number, select one MR from them randomly. 
        \item \label{two} Remove each $f \in F_p$ detected as faulty by that MR from $F_p$.
        \item Repeat steps~\ref{one} and~\ref{two} until all the possible faults are revealed. 
        \end{enumerate}
    \item Select the top \emph{n} MRs from the prioritized ordering to execute based on the resources available for testing.
\end{enumerate}
\subsection{Coverage-based MR Prioritization} \label{Coverage based approach}
Previous studies have shown that MRs with diverse executions in their corresponding source and follow-up test case executions would detect more faults ~\cite{cao2013correlation}. Diverse executions would result in executing more parts of the program under test (i.e. higher coverage) compared to an MR with less diverse executions in its source and follow-up test cases. Based on this intuition we developed the following method for prioritizing MRs based on the statements/branches covered by their corresponding source and follow-up test cases.

 Let the current version of the SUT be $v_k$. Coverage-based MR prioritization uses statement or branch coverage information collected when testing the previous version of the SUT for prioritizing MRs as follows: 
\begin{enumerate}
\item Let the set of source test cases used for testing the version $v_{k-1}$ of the SUT  be the \emph{prioritizing source test cases} $(T_{sp})$.
    \item Let the set of follow-up test cases used for testing the version $v_{k-1}$ of the SUT be the \emph{prioritizing follow-up test cases} $(T_{fp})$. 

\item For each $MR_i$, log the statements (or branches) that were executed when running the corresponding source and follow-up test cases from $(T_{sp})$ and $(T_{fp})$ on version $v_{k-1}$ of the SUT.
\item For each $MR_i$, compute the union of statements (or branches) executed by their source and follow-up cases. 
\item Use the following greedy approach based on the statement (or branch) coverage to create the prioritized ordering of the MRs for testing version $v_k$.
    \begin{enumerate}
       \item Select the MR that covers the highest number of statements (or branches) and place it in the prioritized MR ordering. If there are multiple MRs with the same highest coverage, select one MR from them randomly.\label{step1} 
    \item Remove the statements (or branches) covered by that MR from further consideration.\label{step2} 
    \item Repeat Steps~\ref{step1} and~\ref{step2} until all the possible statements (or branches) are covered. 
    \end{enumerate}
\end{enumerate}

The coverage-based prioritization is explained with an example here. Let P be the target program performing a particular functionality. We define MRs for the program P. The source test case $(T_{sp})$ is obtained by using any traditional test case generation techniques for testing P. The follow-up test cases $(T_{fp})$ is obtained from $(T_{sp})$ by applying MR. The  branches/statements executed are logged while running $(T_{sp})$ and $(T_{fp})$ on the system under test. The greedy algorithm is applied and rank the MRs based on the statements/branches covered.

%% file: Evaluation.tex
\section{Evaluation} \label{section:Eval}
In this work, we plan to evaluate the utility of the developed prioritization approaches on the following aspects: (1) Fault detection effectiveness of MR sets, (2) Effective number of MRs required for testing, and (3) Time taken to detect a fault. To the best of our knowledge,  there are no existing methods for MR prioritization. Therefore, to evaluate the effectiveness of the fault-based and coverage-based prioritizing approaches, we developed (1) a \emph{random baseline}: this represents the current practice of executing source and follow-up test cases of the MRs in random order. (2) an \emph{optimal ordering}: this represents the MR order that yields the maximum achievable fault detection effectiveness with the given set of MRs and source/follow-up test cases.

We conducted experiments to find answers for the following research questions:
\begin{enumerate}
    \item Research Question 1 (RQ1): \emph{Are the proposed MR prioritization approaches more effective than the random baseline?}
    \item Research Question 2 (RQ2): \emph{How do the proposed MR prioritization approaches perform compared to the optimal ordering?}
    \item Research Question 3 (RQ3): \emph{How does the fault-based prioritization approach perform against statement and branch coverage based prioritization approaches?}
    \end{enumerate}

\subsection{Evaluation Procedure}
\label{evalProc}
In order to answer the above research questions, we carry out the following validation procedure:

\begin{enumerate}
\item We used mutants generated using a mutation engine to represent the faults in $F_p$ described in Section~\ref{Mutation-based approach}. $T_{sp}$ and $T_{fp}$ are executed on $F_p$ and any mutants that gave exceptions or mutants that did not finish execution within a reasonable amount of time were removed from this set.  
\item We create a set of source test cases for the SUT independent of the $T_{sp}$ used in Step 1 for validation. These source test cases will be referred to as the \emph{validation source test cases} $(T_{sv})$. Then we create the follow-up test cases according to the MRs used for conducting MT on a given SUT. These will be referred to as the \emph{validation follow-up test cases} ($T_{fv}$). 

\item We generate a set of mutants for the SUT independent of the the $F_p$ mentioned in Section~\ref{Mutation-based approach}. These mutants will be referred to as the \emph{validation set of faults} $(F_v)$.  To generate $F_v$, we use a different mutation engine than the one used for generating $F_p$. Then from the generated mutants, we remove mutants that give exceptions or mutants that do not finish execution within a reasonable amount of time from further consideration. Also, we remove any mutants that are making the same syntactic change as the mutants in $F_p$ from $F_v$. 

\item We use the method described in Section~\ref{Mutation-based approach} to obtain the fault-based MR ordering. Then we applied the obtained MR ordering to $F_v$ and logged the mutant killing information.

\item We use the method described in Section~\ref{Coverage based approach} to obtain the statement-coverage based MR ordering. Then we applied the obtained coverage-based MR ordering to $F_v$ and logged the mutant killing information of the MRs.

\item We used the method described in Section~\ref{Coverage based approach} to obtain the branch-coverage based MR ordering. Then we applied the obtained coverage-based MR ordering to $F_v$ and logged the mutants killing information of the MRs.

\item \textit{Creating the random baseline}: We generate 100 random MR orderings and calculate the average fault detection rate of the 100 random baseline. Then apply each of those orderings to $F_v$. The mutant killing information for each of those random orderings is logged.

\item \textit{Creating the optimal ordering}: Generate by applying the greedy approach described in Section~\ref{Mutation-based approach} to $F_v$ as follows:
\begin{enumerate}
\item Select the MR that kills the largest number of mutants in $F_v$ and place it in the optimal ordering.
\item Remove the mutants killed by that MR from $F_v$.
\item Repeat the previous two steps until all the possible mutants are killed.
\end{enumerate}
\end{enumerate}

\subsection{Evaluation Measures}
\label{evalMeasures}
We used the following measures to evaluate the effectiveness of the MR orderings generated by our MR prioritization approaches:
\begin{enumerate}
    \item To measure the \emph{fault detection effectiveness} of a given set of MRs, we use the percentage of mutants killed by those MRs. 
    \item To calculate the \emph{effective MR set size} we used the following approach: typically, the fault detection effectiveness of  MT  increases  as the number of MRs used for testing increases. However, after a certain number of MRs, the rate of increase in fault detection slows due to factors such as redundancy of MRs. Therefore when there is no significant increase in the fault detection between two MR sets of consecutive sizes of size $m$ and $m+1$, where the MR set of size $m+1$ is created by adding one MR to the MR set set of size $m$, the effective MR set size can be determined. That is, if the difference in fault detection effectiveness of MR set of size $m$ and MR set of size $m+1$ is less than some threshold value, $m$ would be the effective MR set size that should be used for testing. Determining this threshold value should be done considering the critical nature of the SUT. In this work, we used two threshold values of 5\% and 2.5\% as used in previous related work for determining the oracle data set size~\cite{gay2015automated}.
    \item We used the following approach to find the \emph{average time taken to detect a fault}: for each killable mutant $m$ in $F_v$, we computed the time taken to kill the mutant ($t_m$) by computing the time taken to execute the source and follow-up test cases of the MRs in a given prioritized order until $m$ is killed (here it is assumed that the source and follow-up test cases for each MR are executed sequentially, even though in practice it might be possible to execute them in parallel). Then the average time taken to detect a fault is computed using the following formula: $$\frac{\sum t_m}{\#killable\, mutants\, in\, F_v}$$
\end{enumerate}

\subsection{Subject Programs and MRs}
    
   We applied the above-mentioned validation procedure on the following three applications from diverse domains for evaluating our proposed MR prioritization methods. 
    \begin{itemize}
        \item BBMap\footnote{\url{https://jgi.doe.gov/data-and-tools/bbtools/bb-tools-user-guide/BBMap-guide/}}: performs global alignment of RNA and DNA sequence reads. The inputs to the program are a set of reads and a reference genome that is used as the alignment reference. The output is the mapping of reads to the reference genome. 
        \item IBk\footnote{\url{http://weka.sourceforge.net/doc.dev/weka/classifiers/lazy/IBk.html}}: K-nearest neighbours (KNN) classifier implementation  in the Weka machine learning library~\cite{Aha1991}. The input to IBk is a training data set,  and a test data set to represent in .arff format. The output is the classification predictions made on the instances in the test data set. 
        \item LingPipe\footnote{\url{http://alias-i.com/}}: a broad tool kit for processing text using computational linguistics and often used in bioinformatics for bio-entity recognition. In this work, we only use the ``bio-entity recognition'' functionality that focuses on extracting biomedical terms from text and assigning them to specific categories. Input to the bio-entity recognition problem is a set of bio-medical article and it produces extracted bio-entities as the output. 
    \end{itemize}
    
These three applications fall into the category of non-testable programs and represent the programs where MT can be effectively utilized to conduct regression testing.    

\subsection{Metamorphic Relations}
\begin{figure}[h]
\centering
  \begin{subfigure}[b]{0.49\textwidth}
    \includegraphics[width=\textwidth]{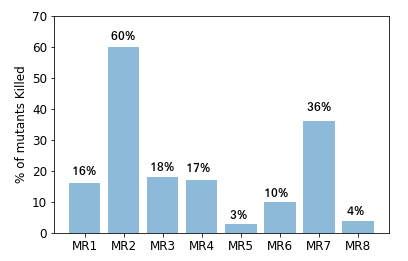}
    \caption{BBMap}
    \label{fig:Killingrate_BBMap}
  \end{subfigure}
  
   ~
     \begin{subfigure}[b]{0.49\textwidth}
    \includegraphics[width=\textwidth]{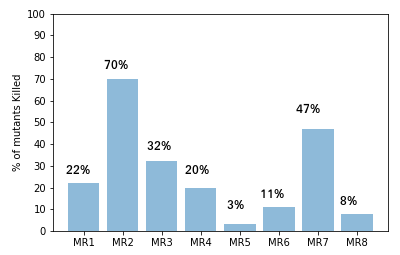}
    \caption{Killing rate of MRs for Major Mutants}
    \label{fig:Killingrate_Major_BBMap}
  \end{subfigure}
  ~
  \begin{subfigure}[b]{0.49\textwidth}
    \includegraphics[width=\textwidth]{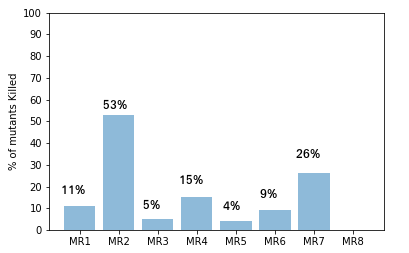}
    \caption{Killing rate of MRs for PIT Mutants}
    \label{fig:Killingrate_PIT_BBMap}
  \end{subfigure}
  \\
   \caption{Fault detection effectiveness of individual MRs for BBMap Programs}
    \label{Fig:mutantkilling_allsubject}
\end{figure}

To conduct MT on the above mentioned subject programs, we used a set of MRs developed in previous studies. For testing BBMap, we used eight MRs (listed in Appendix~\ref{MR BBMap}) developed by Giannoulatou et al. based on the expected behaviour of short-read alignment software~\cite{giannoulatou2014verification}.  All of these MRs specify modifications to the reads supplied as the input, such as by shuffling the reads randomly, duplicating reads, and removing reads. We evaluated the fault detection effectiveness of these MRs on BBMap (see Figure~\ref{fig:Killingrate_BBMap}) using mutation testing (details of mutation testing is provided in Section~\ref{sec:mutants}). As shown in figure~\ref{fig:Killingrate_BBMap}, MR2 (Mapped reads) reported the highest fault detection effectiveness detecting 60\% mutants, followed by MR7 (reverse complement of reads), which detected 36\% mutants. MR5 (quality score increase of reads) had the lowest fault detection effectiveness at 3\%.
Figure~\ref{Fig:mutantkilling_bbmapmutants} shows the killing rate of MRs for Major and PIT mutants. We observe that MR2 performs the best in both the mutant set. Similarly, MR7 performs second best in both the mutant set. The MRs show minimal variation of 1-27\% in the mutant killing rate between the mutant set.

\begin{figure}[h]
\centering
  \begin{subfigure}[b]{0.49\textwidth}
    \includegraphics[width=\textwidth]{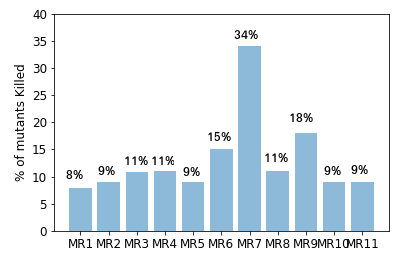}
    \caption{ IBk}
    \label{fig:Killingrate_IBK}
  \end{subfigure}
  \begin{subfigure}[b]{0.49\textwidth}
    \includegraphics[width=\textwidth]{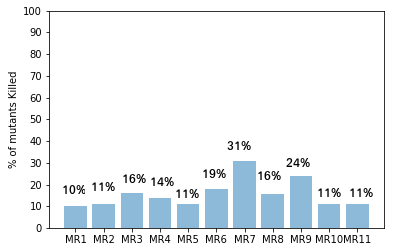}
    \caption{Killing rate of MRs for Major Mutants}
    \label{fig:Killingrate_Major_IBk}
  \end{subfigure}
  ~
  \begin{subfigure}[b]{0.49\textwidth}
    \includegraphics[width=\textwidth]{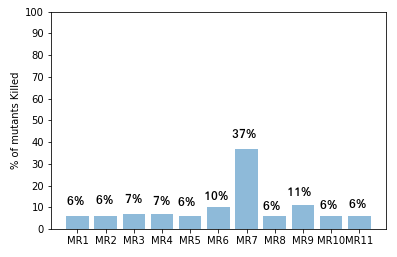}
    \caption{Killing rate of MRs for $\mathbf{\mu}$Java mutants}
    \label{fig:Killingrate_Mujava_IBk}
  \end{subfigure}
   \caption{Fault detection effectiveness of individual MRs for IBK Programs}
    \label{Fig:mutantkilling_allsubject}
\end{figure}

\begin{figure}[h]
\centering
  \begin{subfigure}[b]{0.49\textwidth}
    \includegraphics[width=\textwidth]{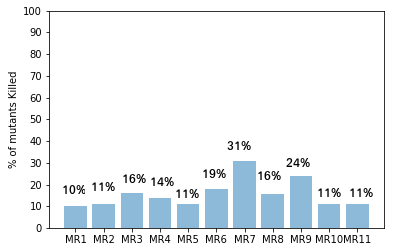}
    \caption{Killing rate of MRs for Major Mutants}
    \label{fig:Killingrate_Major_IBk}
  \end{subfigure}
  ~
  \begin{subfigure}[b]{0.49\textwidth}
    \includegraphics[width=\textwidth]{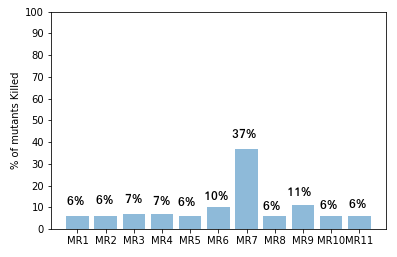}
    \caption{Killing rate of MRs for $\mathbf{\mu}$Java mutants}
    \label{fig:Killingrate_Mujava_IBk}
  \end{subfigure}
  \\
    \caption{Fault detection effectiveness of individual MRs for IBk Program}
    \label{Fig:mutantkilling_IBkmutants}
\end{figure}

For conducting MT on IBk, we used 11 MRs developed by Xie et al.~\cite{xie2011testing} (listed in Appendix~\ref{MR kNN}). These MRs are developed based on the user expectations of supervised classifiers. These MRs modify the training and testing data so that the predictions do not change between the source and follow-up test cases. We evaluated the fault detection effectiveness of these MRs on IBK using mutation testing (see Figure~\ref{fig:Killingrate_IBK}). The result indicated that MR7 (permutation of class labels) provided the highest fault detection rate of 34\% in terms of mutant killing. The lowest fault detection rate of 8\% is provided by MR1 ( consistency with affine transformation), MR2 (permutation of attribute). Figure~\ref{Fig:mutantkilling_IBkmutants} shows the killing rate of MRs for Major and PIT mutants. We observe that MR7 performs the best in both the mutant set. Similarly, MR9 performs second best in both the mutant set. The MR1 to MR6 show minimal variation of 4-9\% and MR7 to MR11 show a variation of 4-13\% in the mutant killing rate between the mutant set.  

\begin{figure}[h]
\centering
  \begin{subfigure}[b]{0.49\textwidth}
    \includegraphics[width=\textwidth]{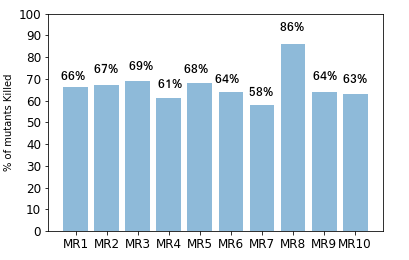}
    \caption{ LingPipe}
    \label{fig:Killingrate_LingPipe}
  \end{subfigure}
  \\
 \begin{subfigure}[b]{0.49\textwidth}
    \includegraphics[width=\textwidth]{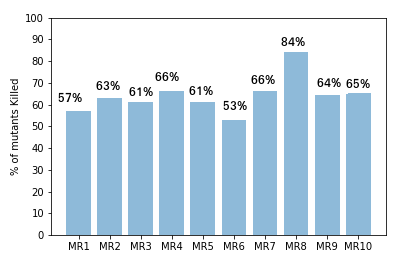}
    \caption{Killing rate of MRs for Major Mutants}
    \label{fig:Killingrate_Major_LingPipe}
  \end{subfigure}
  ~
  \begin{subfigure}[b]{0.49\textwidth}
    \includegraphics[width=\textwidth]{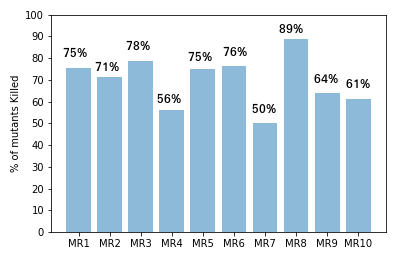}
    \caption{Killing rate of MRs for PIT mutants}
    \label{fig:Killingrate_Mujava_LingPipe}
  \end{subfigure}
  \\
    \caption{Fault detection effectiveness of individual MRs for LingPipe Program}
    \label{Fig:lingpipe_mutant}
\end{figure}

For testing the bio-entity recognition task in LingPipe, we used ten MRs that were developed in our previous work~\cite{srinivasan2018quality}. Appendix~\ref{MR LingPipe} lists these MRs with additional constraints added to them to make them more generalizable. We describe three categories of MRs for testing bio-entity recognition software. These MR categories are addition, deletion, and shuffling. In addition relations, we extend the text by adding new text such that the sentence/paragraph boundaries are preserved. In the deletion relation, text is truncated by removing part of it such that the sentence boundaries are preserved. In the shuffling relations, we shuffle portions of the text. Figure~\ref{fig:Killingrate_LingPipe} shows the fault detection effectiveness (evaluated using mutation testing) of these MRs when used for testing the bio-entity recognition task in LingPipe. MR8 (Removing some words from a list of random word of an article) performed best by providing 86\% overall killing rate. The second best MR was MR3 (adding paragraph to an article) which provides an overall mutant killing rate of 69\%. The least performing MR was MR7 (permutation of class labels) that provided an overall 58\% mutant killing rate. Figure~\ref{Fig:lingpipe_mutant} shows the killing rate of MRs for Major and PIT mutants. We observe that MR8 performs the best in both the mutant set. The MRs show minimal variation of 0-17\% in the mutant killing rate between the mutant set.

\begin{figure}[h]
\centering
  \begin{subfigure}[b]{0.49\textwidth}
    \includegraphics[width=\textwidth]{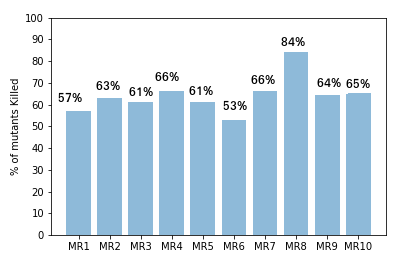}
    \caption{Killing rate of MRs for Major Mutants}
    \label{fig:Killingrate_Major_LingPipe}
  \end{subfigure}
  ~
  \begin{subfigure}[b]{0.49\textwidth}
    \includegraphics[width=\textwidth]{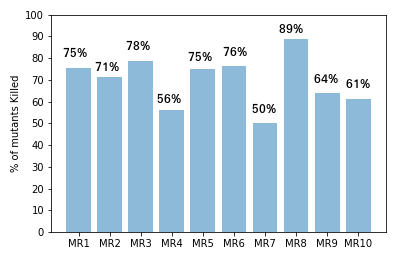}
    \caption{Killing rate of MRs for PIT mutants}
    \label{fig:Killingrate_Mujava_LingPipe}
  \end{subfigure}
  \\
    \caption{Fault detection effectiveness of individual MRs for LingPipe Program}
    \label{Fig:lingpipe_mutant}
\end{figure}

\subsection{Source and Follow-up Test Cases}
\label{testcases}

As we described before, MT involves the process of generating source and follow-up test cases based on the MRs used for testing. BBMap requires a set of reads and a reference genome as inputs. We used the E.coli reference genome and a set of its reads and Yeast reference genome and a set of its reads~\footnote{http://genome.ucsc.edu/} as source test cases. 
E.coli and Yeast are considered as model genomes and are widely used in the bioinformatics domain for conducting tests. To generate the follow-up test cases using these source test cases, we applied the input transformations described in the MRs listed in Appendix~\ref{MR BBMap}. 

IBk uses a training data set to train the k-nearest neighbor classifier, and a test data set is used to evaluate the performance of the trained classifier. Figure~\ref{fig:knnData} in Appendix~\ref{MR kNN} shows an example dataset that is used as a source test case for IBk; the training data set is on the left and the test data set is on the right in the figure. After executing this source test case, the output will be the class labels predicted for the test data set, which is a value from the set \{0,1,2,3,4\} in this example. We generated ten similar datasets randomly, where each training and test data set contained four numerical attributes and a class label. The attribute values are randomly selected within the range [0, 100]. The values of the class label are randomly selected from the set \{0,1,2,3,4,5\}. The size of the training testing data sets ranges within [0, 200]. To generate the follow-up test cases using these source test cases, we applied the input transformations described in the MRs listed in Section~\ref{MR kNN}. 

For creating source test cases for LingPipe, two biomedical articles with the PMCIDs 100320 and 100325 were obtained through PubMed~\footnote{https://www.ncbi.nlm.nih.gov/pubmed}. 
The sentences and paragraphs used as source test cases for the MRs were picked randomly from these two articles. However, for specific MRs related to removing some words from a list of random words (MR8) and shuffling a list of random words (MR10), 15 random articles were selected, and two sets of 500 random words were chosen from the selected articles to generate the source test case. To generate the follow-up test cases using these source test cases, we applied the input transformations described in the MRs listed in Section~\ref{MR LingPipe}.  

\subsection{Mutant Generation}
\label{sec:mutants}
For each subject program, we aimed at developing two independently generated mutant sets to be used as $F_p$ and $F_v$. For this, we used three automated mutation tools: $\mu$Java\footnote{https://github.com/jeffoffutt/muJava}, PIT\footnote{http://pitest.org/} and Major\footnote{http://mutation-testing.org/}. For generating mutants using $\mu$Java, we used all the method level mutation operators~\cite{ma2006mujava}.  With PIT mutation tool~\cite{coles2016pit} and the Major mutation tool~\cite{just2014major} we used all the available mutation operators provided by the tools. All the mutants the were generated using these tools were \emph{first order mutants}, where a mutation operator is used to make a single modification to the source code to generate the mutant.
 
It is typical for mutation tools to generate some \emph{equivalent mutants}; mutants that are syntactically different but functionally equivalent to the original program~\cite{ammann2016introduction}. Therefore, these equivalent mutants always produce the same output as the original program and cannot be detected using testing. Due to the complexity of the subject programs and a large number of mutants generated, it was practically hard to find the equivalent mutants manually. Thus, we removed mutants that were giving the same output as the original program when executed with the above mentioned source test cases.  
Table~\ref{Mutants} shows the number of mutants used for the evaluation after this filtering.

\begin{table}
\centering
\caption{Mutants generated for the Subjects under Study}
 \label{Mutants}
 \begin{tabular}{|>{\centering\arraybackslash}m{1.3cm}|>{\centering\arraybackslash}m{1.3cm}|>{\centering\arraybackslash}m{1.3cm}|>{\centering\arraybackslash}m{1.3cm}|}
\hline
\textbf{Subject}  & \textbf{\# Major} & \textbf{\# PIT} & \textbf{\# $\mathbf{\mu}$Java} \\ \hline
BBMap    & 102              & 100            &   N/A    \\ \hline
IBk      & 100              &     N/A           & 100      \\ \hline
LingPipe & 95               & 96             &    N/A     \\ \hline
\end{tabular}
\end{table}

\begin{table}[h]
 \centering
    \caption{Validation Setup}
    \label{evalSetup}
\begin{tabular}{|l|l|l|l|l|}
\hline
\textbf{Subject} & \textit{$\mathbf{{T_{sp}}}$} &
\textit{$\mathbf{{F_p}}$} &
\textit{$\mathbf{{T_{sv}}}$} &  \textit{$\mathbf{{F_v}}$} \\ \hline
\multirow{4}{*}{BBMap} & Ecoli & Major & Yeast & PIT \\ \cline{2-5} 
 & Yeast & PIT & Ecoli & Major \\ \cline{2-5} 
 & Ecoli & PIT & Yeast & Major \\ \cline{2-5} 
 & Yeast & Major & Ecoli & PIT \\ \hline \hline
\multirow{4}{*}{IBk} & Test1 & Mujava & Test2 & Major \\ \cline{2-5} 
 & Test2 & Major & Test1 & PIT \\ \cline{2-5} 
 & Test2 & Mujava & Test1 & Major \\ \cline{2-5} 
 & Test1 & Major & Test2 & Mujava \\ \hline \hline
\multirow{4}{*}{LingPipe} & ID:100325 & Major  & ID:100320 & PIT \\ \cline{2-5} 
 & ID:100320 & Major & ID:100325 & PIT \\ \cline{2-5} 
 & ID:100320 & PIT & ID:100325 & Major \\ \cline{2-5} 
 & ID:100325 & PIT & ID:100320 & Major \\ \hline
\end{tabular}
\end{table}

\begin{table*}[H]
\centering
\caption{Average relative improvement in fault finding\% using 100\% mutation over reduced mutants for KNN}
 \label{RL_100mutation_reduced_KNN}
\begin{tabular}{|c|c|c|c|}
\hline
MR  & 100\% Mutation Vs 10\% Reduction & 100\% Mutation Vs 25\% Reduction & 100\% Mutation Vs 50\% Reduction \\ \hline
MR1 & 42.58                            & 34.16                            & 60.04                            \\ \hline
MR2 & 13.85                            & 10.58                            & -1.25                            \\ \hline
MR3 & 4.27                             & -1.07                            & 2.90                             \\ \hline
MR4 & 4.20                             & -1.07                            & 2.90                             \\ \hline
MR5 & 4.20                             & -1.07                            & 2.90                             \\ \hline
MR6 & 4.11                             & -2.11                            & 2.42                             \\ \hline
MR7 & 2.65                             & -2.11                            & 2.42                             \\ \hline
MR8 & -0.24                            & -12.07                           & -0.24                            \\ \hline
\end{tabular}
\end{table*}

\begin{table*}[H]
\centering
\caption{Average relative improvement in fault finding\% using 100\% mutation over reduced mutants for LingPipe}
 \label{RL_100mutation_reduced_LingPipe}
\begin{tabular}{|c|c|c|c|}
\hline
MR  & 100\% Mutation Vs 10\% Reduction & 100\% Mutation Vs 25\% Reduction & 100\% Mutation Vs 50\% Reduction \\ \hline
MR1 & 9.52                             & 0                                & 0.27                             \\ \hline
MR2 & 6.40                             & -0.16                            & 0                                \\ \hline
MR3 & 5.05                             & -0.16                            & 0.26                             \\ \hline
MR4 & 2.41                             & -0.16                            & 0.26                             \\ \hline
MR5 & 2.01                             & -0.16                            & 0.26                             \\ \hline
MR6 & 0.80                             & -0.16                            & 0.26                             \\ \hline
MR7 & 0.21                             & -0.05                            & 0                                \\ \hline
MR8 & 0.21                             & 0                                & 0                                \\ \hline
\end{tabular}
\end{table*}

\begin{table*}[H]
\centering
\caption{Average relative improvement in fault finding\% using 100\% mutation over reduced mutants for BBMap}
 \label{RL_100mutation_reduced_BBMap}
\begin{tabular}{|c|c|c|c|}
\hline
MR  & 100\% Mutation Vs 10\% Reduction & 100\% Mutation Vs 25\% Reduction & 100\% Mutation Vs 50\% Reduction \\ \hline
MR1 & 26.94                            & 23.43                            & 7.11                             \\ \hline
MR2 & 12.89                            & 6.56                             & 6.17                             \\ \hline
MR3 & 4.38                             & 1.87                             & 1.49                             \\ \hline
MR4 & -3.88                            & -2.23                            & -2.23                            \\ \hline
MR5 & -0.67                            & 0.35                             & 0.35                             \\ \hline
MR6 & 0.51                             & 0                                & -0.34                            \\ \hline
MR7 & 0                                & 0                                & 0                                \\ \hline
MR8 & 0                                & 0                                & 0                                \\ \hline
\end{tabular}
\end{table*}

%% file: Result.tex
\section{Results and Discussion}
\label{section:Results&Discussion}
\begin{figure}
\centering
  \begin{subfigure}[b]{0.49\textwidth}
    \includegraphics[width=\textwidth,height=6cm]{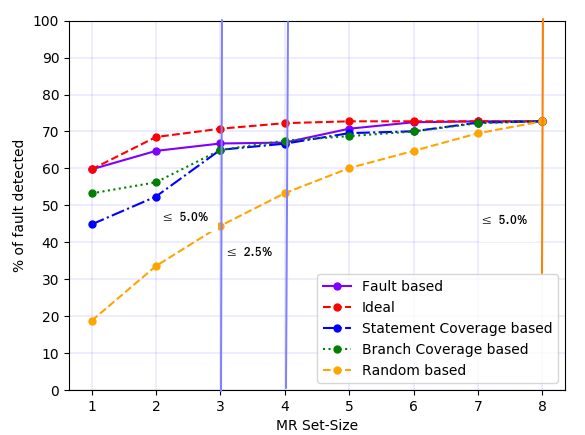}
    \caption{BBMap}
    \label{fig:BBMap_killrateMRs}
  \end{subfigure}
~
  \begin{subfigure}[b]{0.49\textwidth}
    \includegraphics[width=\textwidth,height=6cm]{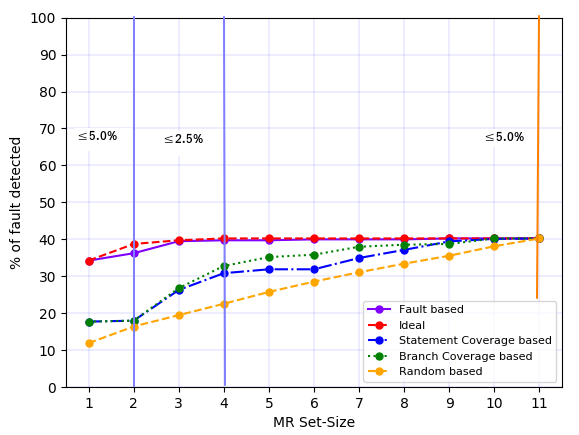}
    \caption{IBk}
    \label{fig:IBK_killrateMRs}
  \end{subfigure}
  \\
  \begin{subfigure}[b]{0.49\textwidth}
      \includegraphics[width=\textwidth,height=6cm]{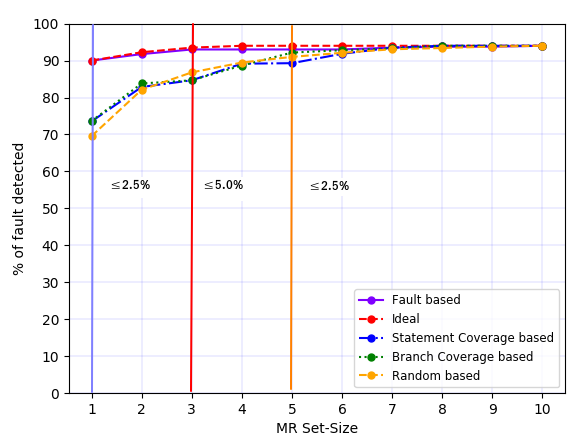}
    \caption{LingPipe}
    \label{fig:LingPipe_killrateMRs}
  \end{subfigure}
  
    \caption{Fault detection effectiveness for BBMap, IBk and LingPipe}
    \label{Fig:allsubject_killrateMRs}
\end{figure}

In this section, we discuss our experimental results and provide answers to the three research questions that we listed in Section~\ref{section:Eval}. For each subject program, we carried out the validation procedure described in Section~\ref{evalProc} using the setup described in Table~\ref{evalSetup}. In this setup, we used the generated mutant sets and the source test cases as $F_p, T_{sp}, F_v,$ and ${T_{sv}}$ in four different configurations. For example, the first evaluation run for BBMap was conducted by executing Ecoli as $T_{sp}$, Major mutants as $F_p$, Yeast as $T_{sv}$, and PIT mutants as $F_{v}$ (refers to row 1 of Table~\ref{evalSetup}). The second evaluation run for BBMap was conducted by executing Yeast as $T_{sp}$, PIT mutants as $F_p$, Ecoli as $T_{sv}$, and Major mutants as $F_{v}$ (refers to row 2 of Table~\ref{evalSetup}). Similarly, two additional evaluation runs were conducted for BBMap using the configurations in rows three and four in Table~\ref{evalSetup}. In Table~\ref{evalSetup} for IBk, Test1 refers to 5 of the randomly generated datasets that we described in Section~\ref{testcases} and Test2 refers to the other five datasets. Figure~\ref{Fig:allsubject_killrateMRs} shows the average fault detection effectiveness for evaluation runs described above vs. the MR set size used for testing, for each subject program. We also plot the percentage of faults detected by the random baseline and the optimal MR ordering for comparison.

\subsection{\textbf{RQ1: Comparison of the MR Prioritization approaches against the Random Baseline} }

\subsubsection{Fault detection effectiveness}

We formulated the following statistical hypothesis to answer RQ1 in the context of fault detection effectiveness:
\begin{itemize}
\item [$H_{1}$:]For a given MR set of size \textit{m}, the fault detection effectiveness of the MR set produced by fault-based approach is higher than that of the random baseline.
\item[$H_{2}$:]For a given MR set of size \textit{m}, the fault detection effectiveness of the MR set produced by statement coverage based approach is higher than that of the random baseline.
\item[$H_{3}$:]For a given MR set of size \textit{m}, the fault detection effectiveness of the MR set produced by the branch coverage based approach is higher than that of the random baseline.
\end{itemize}
The null hypothesis $H_{0x}$ for each of the above defined hypothesis $H_{X}$ is that the fault-based, statement coverage and branch coverage-based approaches perform equal to or worse than the random baseline.

In Table~\ref{tab:RL_methods_random_BBMap}, ~\ref{tab:RL_methods_random_IBk} and ~\ref{tab:RL_methods_random_LingPipe} we list the relative improvement in average fault detection effectiveness for the three MR prioritization methods over the currently used random approach for BBMap, IBk, and LingPipe, respectively. To evaluate the above hypotheses, we used the paired permutation test since it does not make any assumptions about the underlying distribution of the data~\cite{good2013permutation}.  We apply the paired permutation test to each MR set size for each of the subject programs with $\alpha$= 0.05, and the relative improvements that are significant are marked with a *. 
As shown in Table~\ref{tab:RL_methods_random_BBMap}, for BBMap, the three prioritization approaches showed significant improvements in average fault detection effectiveness over the random approach for all the MR set sizes except for the last set size. Particularly, among the different MR set sizes, the increase in fault detection percentage varies from 3.91\% to 218.24\% between the three methods. For example, in a situation where only three MRs can be used for testing BBMap, using a MR prioritization would yield an increase of 46\% - 50\% in fault detection effectiveness. Similar observations can be made for IBk as shown in Table~\ref{tab:RL_methods_random_IBk}, where significant relative improvements in fault detection percentages can be observed for the first 10 out of 11 MR sets. With IBk, the improvement in fault detection effectiveness varies from 5.52\% to 187.12\% among the prioritization methods for different MR set sizes. Therefore, we reject the null hypotheses $H_{01}$, $H_{02}$, and $H_{03}$ for BBMap and IBk. The random and code coverage based chose a highly effective MR at MR set size two. As a result, the relative improvement between the approaches dropped significantly. However, in MR set size three, statement coverage chose a more effective MR than random approach, hence, the increase in relative improvement in MT set size 3.

Table~\ref{tab:RL_methods_random_LingPipe} shows the relative improvement in the average fault detection percentage for LingPipe. While  fault-based MR prioritization shows significant improvements for MR set sizes $m=1$ to $m=7$, statement and branch coverage-based prioritization did not show such consistent improvements across different MR set sizes. Also overall, the relative improvement across MR prioritization methods varies between 0.23\% - 29.4\%, which is less than the other two subject programs. Further, for some MR sets, a random baseline slightly outperformed the statement and branch coverage-based prioritization. Therefore we can only reject $H_{01}$ for MR set sizes  $m=1$ to $m=7$, and we cannot reject $H_{02}$ and $H_{03}$ in general for LingPipe. 

This difference in the performance of LingPipe can be explained using the difference in the fault detection effectiveness of individual MRs for LingPipe compared to the other two subject programs. Figure~\ref{Fig:mutantkilling_allsubject} shows the percentage of mutants killed by the MRs used for testing each subject program. As shown in Figure~\ref{fig:Killingrate_LingPipe}, MRs used for testing LingPipe has a higher average killing rate (66.87\%) with a low variance among the killing rates (standard deviation = 7.71). For MRs used for testing BBMap (see Figure~\ref{fig:Killingrate_BBMap}), the average killing rate is 20.79\% with a standard deviation of 18.93\% and for IBk (see Figure~\ref{fig:Killingrate_IBK}), the average killing rate is 12.93\% with a standard deviation of 7.69\%. Since the variation in killing rates of the MRs was comparatively high in BBMap and IBk, they were able to gain a higher benefit by utilizing MR prioritization compared to LingPipe. Further, except for MR8, the statement and branch coverage of other MRs were similar for LingPipe. Therefore, coverage-based prioritization would yield in a similar performance as the random baseline since the MRs are picked randomly when the coverage is the same. 

\begin{table}[h]
\centering
 \caption{Relative improvement in fault detection effectiveness of the three MR prioritization approaches compared to random baseline for BBMap}
     \label{tab:RL_methods_random_BBMap}
\begin{tabular}{|c|c|c|c|}
\hline
\textbf{MR Set-Size} & \textbf{Fault-based} & \textbf{Statement Cov-based} & \textbf{Branch Cov-based} \\ \hline
1                                     & 218.24\%*               & 138.88\%*                    & 183.62\%*                 \\ \hline
2                                     & 92.70\%*                & 55.95\%*                     & 67.41\%*                  \\ \hline
3                                     & 50.08\%*                & 46.26\%*                     & 46.14\%*                  \\ \hline
4                                     & 25.70\%*                & 25.14\%*                     & 26.64\%*                  \\ \hline
5                                     & 17.72\%*                & 15.72\%*                     & 14.39\%*                  \\ \hline
6                                     & 11.96\%*                & 8.18\%*                      & 8.10\%*                   \\ \hline
7                                     & 4.63\%*                 & 4.13\%*                      & 3.91\%*                   \\ \hline
8                                     & 0\%*                    & 0\%                          & 0\%                       \\ \hline
\end{tabular}
\end{table}

\begin{table}[h]
\centering
 \caption{Relative improvement in fault detection effectiveness of the three MR prioritization approaches compared to random baseline for IBk}
     \label{tab:RL_methods_random_IBk}
\begin{tabular}{|c|c|c|c|}
\hline
\textbf{MR Set-Size} & \textbf{Fault-based} & \textbf{Statement Cov-based} & \textbf{Branch Cov-based} \\ \hline
1                                     & 187.12\%*               & 48.80\%*                     & 48.80\%*                  \\ \hline
2                                     & 120.65\%*               & 9.56\%*                      & 9.56\%*                   \\ \hline
3                                     & 102.65\%*               & 34.87\%*                     & 37.43\%*                  \\ \hline
4                                     & 76.10\%*                & 36.67\%*                     & 45.31\%*                  \\ \hline
5                                     & 54.15\%*                & 23.71\%*                     & 36.50\%*                  \\ \hline
6                                     & 40\%*                   & 11.65\%*                     & 25.47\%*                  \\ \hline
7                                     & 28.58\%*                & 12.39\%*                     & 22.15\%*                  \\ \hline
8                                     & 19.65\%*                & 11.13\%*                     & 15.32\%                   \\ \hline
9                                     & 13.15\%*                & 10.90\%*                     & 8.93\%*                   \\ \hline
10                                    & 5.52\%*                 & 5.22\%*                      & 5.52\%*                   \\ \hline
11                                    & 0\%                     & 0\%                          & 0\%                       \\ \hline
\end{tabular}
\end{table}

\begin{table}[h]
\centering
 \caption{Relative improvement in fault detection effectiveness of the three MR prioritization approaches compared to random baseline for LingPipe}
     \label{tab:RL_methods_random_LingPipe}
\begin{tabular}{|c|c|c|c|}
\hline
\textbf{MR Set-Size} & \textbf{Fault-based} & \textbf{Statement Cov-based} & \textbf{Branch Cov-based}
\\ \hline
1                                     & 29.40\%*                 & 5.73\%*                      & 5.88\%*                   \\ \hline
2                                     & 11.75\%*                & 0.91\%                       & 2.07\%                    \\ \hline
3                                     & 7.14\%*                 & -2.45\%*                    & -2.45\%*                  \\ \hline
4                                     & 3.93\%*                 & -0.34\%                      & -0.90\%*                  \\ \hline
5                                     & 2.19\%*                 & -1.87\%*                     & 1.28\%*                   \\ \hline
6                                     & 0.97\%*                  & -0.28\%                      & 0.76\%*                   \\ \hline
7                                     & 0.48\%*                & 0.48\%                       & 0.24\%                    \\ \hline
8                                     & 0.37\%                  & 0.64\%*                       & 0.64\%*                   \\ \hline
9                                     & -0.02\%                 & 0.23\%*                     & 0.23\%*                   \\ \hline
10                                    & 0\%                     & 0\%                          & 0\%                       \\ \hline
\end{tabular}
\end{table}

\subsubsection{Effective number of MRs used for Testing}
In Figures~\ref{fig:BBMap_killrateMRs},~\ref{fig:IBK_killrateMRs}, and~\ref{fig:LingPipe_killrateMRs}, we show the effective number of MRs using the fault detection thresholds that we described in Section~\ref{evalMeasures}, for BBMap, IBk, and LingPipe, respectively. The purple vertical lines represent effective MR set size when using the fault-based prioritization and the orange vertical lines represent the same for the random baseline. The respective thresholds are shown near each vertical line.

As shown in Figure~\ref{fig:BBMap_killrateMRs}, for BBMap, the effective MR set size is 3, and 4 for the 5\% and 2.5\% thresholds, respectively when fault-based prioritization is used. With the random baseline, the effective MR set size is 8 at the 5\% threshold and the 2.5\% threshold is never met. For IBk, (see Figure~\ref{fig:IBK_killrateMRs}), a similar observation can be made where the effective MR set size with fault-based prioritization is 2 and 4 for the 5\% and 2.5\% thresholds. For the random baseline, effective MR set size is 11. For LingPipe, the effective MR set size is one for the 2.5\% threshold with fault-based prioritization. With the random baseline, 5\% and 2.5\% thresholds are only met with MR set sizes 3 and 5 respectively. Therefore for all the three subject programs, using fault-based prioritization resulted in a considerable reduction in the effective MR set size compared to using a random ordering of MRs. In practice, this reduction would translate into saving resources used for testing.

\subsubsection{The average time taken to detect a fault } Tables ~\ref{time_mutation_random_bbmap} and~\ref{time_mutation_random_LingPipe} shows the average time taken to detect a fault in BBMap and LingPipe respectively, for the four validation configurations described in Table~\ref{evalSetup}. As shown in these results, using fault-based prioritization resulted in significant reductions in the average time taken to detect a fault ranging from 23\%-61\% relative to the random baseline for BBMap and 10\%-40\% relative to the random baseline for LingPipe.
\begin{table}[h]
\centering
\caption{Comparison of average time taken to detect a fault using fault-based MR prioritization and Random baseline for BBMap}
 \label{time_mutation_random_bbmap}
 \begin{tabular}{|p{2cm}|p{2cm}|p{2cm}|p{2cm}|p{2cm}|}
\hline

$\mathbf{{T_{sv}}}$ &\textbf{$\mathbf{{F_v}}$} & \textbf{Fault-based} & \textbf{Random baseline}&\textbf{\% Time reduction} \\ \hline
 Ecoli &    PIT & 141s  & 244s & 42\%   \\ \hline
Ecoli & Major  & 4704s & 8972s & 48\%   \\ \hline
Yeast &  Major & 8568s  & 11140s & 23\% \\ \hline
 Ecoli & PIT & 58s & 149s & 61\% \\ \hline
\end{tabular}
\end{table}

\begin{table}[h]
\centering
\caption{Comparison of average time taken to detect a fault using fault-based MR prioritization and Random baseline for LingPipe}
 \label{time_mutation_random_LingPipe}
 \begin{tabular}{|p{2cm}|p{2cm}|p{2cm}|p{2cm}|p{2cm}|}
\hline
$\mathbf{{T_{sv}}}$ &\textbf{$\mathbf{{F_v}}$} & \textbf{Fault-based} & \textbf{Random baseline}&\textbf{\% Time reduction} \\ \hline
 ID:100325 &    PIT & 76s  & 94s & 19\%   \\ \hline
 ID:100320 & Major  & 125s & 210s & 40\%   \\ \hline
 ID:100320 &  PIT & 96s  & 160s & 40\% \\ \hline
  ID:100325 & Major & 113s & 126s & 10\% \\ \hline
\end{tabular}
\end{table}

\subsection{RQ2: Comparison of proposed MR Prioritization approaches against Optimal Ranking}

\begin{table}[h]
\centering
\caption{Average relative improvement in fault detection effectiveness of Optimal approach over fault-based, Statement and Branch coverage-based approach for BBMap}
\label{tab:Optimal_Methods_BBMap}
\begin{tabular}{|c|c|c|c|}
\hline
\textbf{MR Set-Size} & \textbf{Fault-based} & \textbf{Statement Cov-based} & \textbf{Branch Cov-based} \\ \hline
1                                     & 0\%                                          & 33.22\%*                                          & 12.20\%*                                       \\ \hline
2                                     & 5.79\%*                                      & 30.70\%*                                          & 21.77\%*                                       \\ \hline
3                                     & 5.99\%*                                      & 8.76\%*                                           & 8.84\%*                                        \\ \hline
4                                     & 7.83\%*                                      & 8.32\%*                                           & 7.03\%*                                        \\ \hline
5                                     & 2.82\%*                                      & 4.60\%*                                           & 5.81\%*                                        \\ \hline
6                                     & 0.34\%*                                      & 3.85\%*                                           & 3.92\%*                                        \\ \hline
7                                     & 0\%                                          & 0.48\%*                                           & 0.69\%*                                        \\ \hline
8                                     & 0\%                                          & 0\%                                               & 0\%                                            \\ \hline
\end{tabular}
\end{table}

\begin{table}[h]
\centering
\caption{Average relative improvement in fault detection effectiveness of Optimal approach over fault-based, Statement and Branch coverage-based approach for IBk}
\label{tab:Optimal_Methods_IBk}
\begin{tabular}{|c|c|c|c|}
\hline
\textbf{MR Set-Size} & \textbf{Fault-based} & \textbf{Statement Cov-based} & \textbf{Branch Cov-based} \\ \hline
1                                     & 0\%                     & 92.95\%*                     & 92.95\%*                  \\ \hline
2                                     & 6.89\%*                 & 115.27\%*                    & 115.27\%*                 \\ \hline
3                                     & 0.63\%*                 & 51.14\%*                     & 48.32\%*                  \\ \hline
4                                     & 1.25\%*                 & 30.47\%*                     & 22.71\%*                  \\ \hline
5                                     & 1.25\%*                 & 26.17\%*                     & 14.34\%*                  \\ \hline
6                                     & 0.62\%*                 & 26.17\%*                     & 12.27\%*                  \\ \hline
7                                     & 0.62\%                  & 15.12\%*                     & 5.92\%*                   \\ \hline
8                                     & 0.62\%                  & 8.34\%*                      & 4.40\%*                   \\ \hline
9                                     & 0\%                     & 2.02\%*                      & 3.87\%*                   \\ \hline
10                                    & 0\%                     & 0.28\%*                      & 0\%                       \\ \hline
11                                    & 0\%                     & 0\%                          & 0\%                       \\ \hline
\end{tabular}
\end{table}

\begin{table}[h]
\centering
\caption{Average relative improvement in fault detection effectiveness of Optimal approach over fault-based, Statement and Branch coverage-based approach for LingPipe}
\label{tab:Optimal_Methods_LingPipe}
\begin{tabular}{|c|c|c|c|}
\hline
\textbf{MR Set-Size} & \textbf{Fault-based} & \textbf{Statement Cov-based} & \textbf{Branch Cov-based} \\ \hline
1                                     & 0\%                     & 22.37\%*                     & 22.21\%*                  \\ \hline
2                                     & 0.54\%*                 & 11.34\%*                     & 10.08\%*                  \\ \hline
3                                     & 0.53\%*                 & 10.43\%*                     & 10.43\%*                  \\ \hline
4                                     & 1.07\%*                 & 5.42\%*                      & 6.01\%*                   \\ \hline
5                                     & 1.07\%*                 & 5.27\%*                      & 1.98\%*                   \\ \hline
6                                     & 1.07\%*                 & 2.35\%*                      & 1.28\%*                   \\ \hline
7                                     & 0.53\%*                 & 0.53\%*                      & 0.77\%*                   \\ \hline
8                                     & 0.26\%*                 & 0\%                          & 0\%                       \\ \hline
9                                     & 0.26\%*                 & 0\%                          & 0\%                       \\ \hline
10                                    & 0\%                     & 0\%                          & 0\%                       \\ \hline
\end{tabular}
\end{table}

In order to answer the second research question, we formulated the following statistical hypothesis:

\begin{itemize}
    \item[$H_{4}$:]For a given MR set of size $m$, the fault detection effectiveness of the MR set produced using Optimal ordering is higher than the fault detection effectiveness of the MR set produced by the fault-based prioritization. 
    
    \item[$H_{5}$:] For a given MR set of size $m$, the fault detection effectiveness of the MR set produced using Optimal ordering is higher than the fault detection effectiveness of the MR set produced by the statement coverage-based prioritization.
    
    \item[$H_{6}$:]For a given MR set of size $m$, the fault detection effectiveness of the MR set produced using Optimal ordering is higher than the fault detection effectiveness of the MR set produced by the branch coverage-based prioritization.
    \end{itemize}

The null hypothesis $H_{0x}$ for each of the above defined Hypotheses $H_{x}$ is that the MR sets generated by the corresponding MR prioritization approach performs equal to or worse than the MR sets generated by the optimal ordering in terms of fault detection effectiveness. Table~\ref{tab:Optimal_Methods_BBMap}, ~\ref{tab:Optimal_Methods_IBk} and ~\ref{tab:Optimal_Methods_LingPipe} show the relative improvement in fault detection effectiveness between the optimal ordering and the MR prioritization methods developed by us. The results that are significant at $\alpha=0.05$ are shown with a \textasteriskcentered. As discussed above, Optimal ordering represents the best ordering of MRs if the faults are known in advance. Thus, it represents the upper bound of fault detection for a given set of MRs and source/follow-up test cases. Here we attempt to find how closely do our MR prioritization approach performs compared to this upper bound.  

As shown in Table~\ref{tab:Optimal_Methods_BBMap},~\ref{tab:Optimal_Methods_IBk},~\ref{tab:Optimal_Methods_LingPipe} fault-based approach reported a 0\% increase for the MR set of size $m=1$. This indicates that the fault-based MR prioritization is able to identify the most effective MR from the given set of MRs and placed it first in the prioritized MR set for all the three subject programs. However, the coverage-based approaches were not able to achieve this. In particular, for IBk the first two MR sets generated using the coverage-based MR prioritization approaches reported a significantly lower fault detection rate when compared to the Optimal ordering. For IBk, MR7 reported the highest fault detection effectiveness. However, this MR did not report the highest statement or branch coverage. Thus, coverage-based MR prioritization approaches failed to choose this MR first in the generated prioritized order. This is not surprising since IBk is a supervised machine learning classifier that creates a machine learning model form the training data. For such programs only achieving high coverage in the source code is not enough for achieving a high fault detection effectiveness, since that would not essentially represent the ability to identify issues in the developed machine learning model. Therefore, for those types of programs, it would be more effective to use fault-based MR prioritization that directly uses the fault detection capabilities of MRs for creating the prioritized MR ordering. 

In general, the results show that while the MR sets produced by the Optimal ordering have comparatively higher fault detection percentages than the three prioritization methods, the gap between them is considerably small, in particular for the fault-based prioritization approach. Therefore, we can conclude that the fault-based MR prioritization is quite effective in producing a MR ordering that is comparably effective as the Optimal ordering.

\subsection{RQ3: Comparison of the fault-based MR prioritization approach against the coverage-based prioritization approaches}

\begin{table*}[h]
\centering
\caption{Average relative improvement in fault detection effectiveness using fault-based approach over Statement and Branch coverage-based prioritization approaches}
\label{tab:RL_mutation_cov_subjects}
\begin{tabular}{|>{\centering\arraybackslash}m{1.75cm}||>{\centering\arraybackslash}m{1.15cm}|>{\centering\arraybackslash}m{1.15cm}||>{\centering\arraybackslash}m{1.75cm}|>{\centering\arraybackslash}m{1.75cm}||>{\centering\arraybackslash}m{1cm}>{\centering\arraybackslash}m{1.75cm}}
\hline
\multicolumn{1}{|m{1.05cm}||}{\multirow{2}{*}{\textbf{MR Set-Size}}} & \multicolumn{2}{c||}{\textbf{BBMap}}                                           & \multicolumn{2}{c||}{\textbf{IBk}}                        & \multicolumn{2}{c|}{\textbf{LingPipe}}                                                             \\ \cline{2-7} 
\multicolumn{1}{|c||}{}                                      & \multicolumn{1}{>{\centering\arraybackslash}m{1.15cm}|}{\textbf{Statement Cov-based}} & \textbf{Branch Cov-based} & \textbf{Statement Cov-based} & \textbf{Branch Cov-based} & \multicolumn{1}{>{\centering\arraybackslash}m{1.75cm}|}{\textbf{Statement Cov-based}} & \multicolumn{1}{>{\centering\arraybackslash}m{1.75cm}|}{\textbf{Branch Cov-based}} \\ \hline
1                                                         & \multicolumn{1}{c|}{33.22\%*}                     & 12.20\%*                  & 92.95\%*                     & 92.95\%*                  & \multicolumn{1}{c|}{23.37\%*}                     & \multicolumn{1}{c|}{22.21\%*}                  \\ \hline
2                                                         & \multicolumn{1}{c|}{23.56\%*}                     & 15.11\%*                  & 101.38\%*                    & 101.38\%*                 & \multicolumn{1}{c|}{10.74\%*}                     & \multicolumn{1}{c|}{9.48\%*}                   \\ \hline
3                                                         & \multicolumn{1}{c|}{2.61\%*}                      & 2.69\%*                   & 50.19\%*                     & 47.38\%*                  & \multicolumn{1}{c|}{9.84\%*}                      & \multicolumn{1}{c|}{9.84\%*}                   \\ \hline
4                                                         & \multicolumn{1}{c|}{0.44\%}                       & 0.74\%                    & 28.84\%*                     & 21.18\%*                  & \multicolumn{1}{c|}{4.29\%*}                      & \multicolumn{1}{c|}{4.88\%*}                   \\ \hline
5                                                         & \multicolumn{1}{c|}{1.72\%*}                      & 2.90\%*                   & 24.60\%*                     & 12.92\%*                  & \multicolumn{1}{c|}{4.15\%*}                      & \multicolumn{1}{c|}{0.90\%*}                   \\ \hline
6                                                         & \multicolumn{1}{c|}{3.49\%*}                      & 3.57\%*                   & 25.39\%*                     & 11.57\%*                  & \multicolumn{1}{c|}{1.27\%*}                      & \multicolumn{1}{c|}{0.20\%*}                   \\ \hline
7                                                         & \multicolumn{1}{c|}{0.48\%*}                      & 0.69\%*                   & 14.40\%*                     & 5.26\%                    & \multicolumn{1}{c|}{0\%}                          & \multicolumn{1}{c|}{0.24\%}                    \\ \hline
8                                                         & \multicolumn{1}{c|}{0\%}                          & 0\%                       & 7.67\%*                      & 3.76\%*                   & \multicolumn{1}{c|}{-0.26\%*}                     & \multicolumn{1}{c|}{-0.26\%}                   \\ \hline
\multicolumn{1}{|c||}{9}                                   & \multicolumn{1}{c}{}                              & \multicolumn{1}{c|}{}     & \multicolumn{1}{c|}{2.02\%*} & 3.87\%*                   & \multicolumn{1}{c|}{-0.26\%*}                     & \multicolumn{1}{c|}{-0.26\%}                   \\ \cline{1-1} \cline{4-7} 
\multicolumn{1}{|c||}{10}                                  & \multicolumn{1}{c}{}                              & \multicolumn{1}{c|}{}     & \multicolumn{1}{c|}{0.28\%}  & 0\%                       & \multicolumn{1}{c|}{0\%}                          & \multicolumn{1}{c|}{0\%}                       \\ \cline{1-1} \cline{4-7} 
\multicolumn{1}{|c||}{11}                                  & \multicolumn{1}{c}{}                              & \multicolumn{1}{c|}{}     & \multicolumn{1}{c|}{0\%}     & 0\%                       &                                                   &                                                \\ \cline{1-1} \cline{4-5}
\end{tabular}
\end{table*}

In this research question, we evaluate whether the fault-based ranking approach outperforms statement and branch coverage-based approaches. In order to answer the research question, we formulate the following hypotheses:

\begin{itemize}
    \item[$H_{7}$:] For a given MR set of size $m$, the fault detection effectiveness of the MR set produced using fault-based prioritization is higher than the fault detection effectiveness of the MR set produced by the statement coverage-based prioritization.
    \item[$H_{8}$:] For a given MR set of size $m$, the fault detection effectiveness of the MR set produced using fault-based prioritization is higher than the fault detection effectiveness of the MR set produced by the branch coverage-based prioritization.
\end{itemize}

The null hypothesis $H_{0x}$ for each of the above defined Hypotheses $H_{x}$ is that the MR sets generated by the fault-based prioritization method performs equal to or worse than the MR sets generated by the statement and branch coverage-based prioritization in terms of fault detection effectiveness.

Table~\ref{tab:RL_mutation_cov_subjects} shows the relative improvement in fault detection percentage between the MR sets generated by fault-based prioritization approach and the MR sets generated by a statement/branch coverage based prioritization approaches for BBmap, IBk and LingPipe, respectively. For all the three programs, the majority of MR sets generated by fault-based prioritization had significantly higher fault detection effectiveness compared to the MR sets generated by statement and branch coverage based prioritization. Particularly, for IBk there are improvements greater than 10\% for the MR sets of sizes $m=1$ to $m=7$. Therefore, in general,  we can reject the null hypotheses $H_{07}$ and $H_{08}$.

In practice, these increases in improvements in fault detection effectiveness translate to potentially detecting more faults using fault-based MR prioritization compared to coverage-based MR prioritization methods, when there are restrictions on the number of MRs that could be executed due to time/budget constraints. However, it is important to note that fault-based MR prioritization involves a higher overhead compared to coverage-based MR prioritization methods since it requires the MRs to be executed on a set of mutants. Therefore this additional cost needs to be balanced out with the reported benefits above. Thus, we recommend using the fault-based MR prioritization when only a limited number of MRs can be executed due to time/budget constraints.

%% file: ThreattoValidity.tex
\section{Threat to Validity}
\label{section:ThreattoValidity}

\textbf{External Validity: } We only used three subject programs for our validation. However, these three applications are complex systems that are from different domains. We believe that they are good representations of applications that face the oracle problem for which MT is effective and widely used. Thus, we believe that our results are generalizable to applications for which MT is applicable. We have generated approximately 200 mutants for each subject, where approximately 100 mutants used for generating the prioritized MR orderings and approximately 100 mutants are used for validation. These numbers were determined after considering the time and cost taken for executing them. However, it is possible that the number of mutants used is low. 

\textbf{Internal Validity: } For BBMap and LingPipe, we used existing test cases as source test cases. For IBk, source test cases were randomly generated. Generating source test cases using structural coverage criteria such as MC/DC (Modified Condition/ Decision coverage), Decision coverage based test cases might influence the fault detection ability of the MRs and hence result in a different MR ordering. The greedy approach to create prioritized MR ordering selects a MR randomly when multiple MRs detect the same number of faults. This could impact the performance of the approaches. We plan to handle the limitation in the future work. The approaches when used in the scenario of regression testing, prioritized MR order could become invalid when future version undergoes major changes in the functionality of the system. We believe that the invalidity of MR is unlikely to happen since the MRs are necessary properties of the target function or algorithm~\cite{chen2018metamorphic}. Therefore, developers might add more MRs to the existing MRs in the future versions than destroying the existing MRs.

\textbf{Construct Validity: }We measure the fault finding effectiveness of the MRs using mutants rather than real faults. MR prioritization carried out based on real faults might lead to a different result. However, studies conducted out by Just et al.~\cite{just2014mutants} have shown a significant correlation between the detection of seeded faults and the detection of real faults.

%% file: Conclusion.tex
\section{Conclusions}
\label{Section:Conclusion}
In this work, we developed automated approaches for MR prioritization in order to improve the efficiency and effectiveness of MT. These approaches use (1) fault detection information, and (2) statement/branch coverage information to prioritize MRs. We conducted an empirical study to evaluate the effectiveness of the developed approaches using three complex open-source software systems.


The experimental results show that utilizing MR prioritization would increase the fault detection effectiveness of a given MR set and these increases could range up to 218.24\%. In particular, using fault-based MR prioritization reduced the number of MRs that needs to be executed during testing and also reduced the average time taken to detect a fault compared to the currently used method of executing the source and follow-up test cases of the MRs in random order. Further, our results show that fault-based MR prioritization outperforms the coverage-based MR prioritization and performs comparably to the optimal MR ordering. Based on these results, we recommend utilizing MR prioritization to improve the efficiency and effectiveness of MT.

%% file: appendix.tex
\newpage
\section{MRs used for testing the subject programs}
In this section, we discuss the MRs used in each of the subject programs. 
\subsection{MRs for BBMap Program}
\label{MR BBMap}
Below is the set of MRs for testing BBMap program, and the MRs were obtained from Giannoulatou et al.~\cite{giannoulatou2014verification}.
\begin{itemize}
    \item MR1 (Read Addition): The reads in the initial test are duplicated, making the input for the follow-up test twice as large. We also expect the coverage in the follow-up test to be twice of the original coverage. The MR include one source test case (reads) and one follow-up test case (duplicated reads).
    \item MR2 (Mapped Reads): After initial mapping, only the mapped reads are selected and remapped against the reference genome. We expect that all of the reads will be mapped. The MR include one source test case(reads) and one follow-up test case (mapped reads).
    \item MR3 (Correction of errors): After initial mapping, the mapped reads are selected and any mismatch or error is corrected in the reads. We expect the output mapping to remain the same. The MR include one source test case(mapped read) and one follow-up test case( corrected read).
    \item MR4 (Random permutation of reads): The reads in the FASTQ files are reshuffled. The permutation is the same for Read1 and Read2 reads. We expect the output mapping to be the same as the original output. The MR include one source test case (reads) and one follow-up test case (permuted read).
    \item MR5 (Quality score increase of reads): After initial mapping, the quality score for all mapped sequences is increased. We expect the output mapping to remain the same. The MR contain one source test case (reads) and one follow-up test case(quality score increased).
    \item MR6 (Read Removal): Half of the reads from the initial input collection are deleted. We also expect the coverage in the follow-up test to be half of the original coverage. The MR contain one source test case(reads) and one follow-up test case(reads removed).
    \item MR7 (Reverse Complement of reads): The reads are reverse complemented and the corresponding quality values are reversed. We expect the output mapping to be same as the original output. The MR contains one source test case(Reads) and one follow-up test case(reverse complemented reads).
    \item MR8 (Unmapped Reads): After initial mapping, only the unmapped reads are selected and remapped against the reference genome. We expect that none of the reads will be mapped. The MR include one source test case(mapped read) and one follow-up test case(unmapped read).
\end{itemize}

\subsection{MRs for IBK Program}
The MRs used for testing IBK program is discussed below and the MRs were obtained from Xie et al.~\cite{xie2011testing}
\label{MR kNN}
\begin{itemize}
\item MR1 (Consistence with affine transformation): The result should be same, if we apply some affine transformation function, f(x) = kx + b,(k$\neq$ 0), to every value x in some subset of features in the training and testing data. The appropriate value of K=3 (i.e.number of nearest neighbours) such that MR becomes a necessary property for IBK. The MR contain one source test case (training and testing set). The MR contains one source and follow-up test case.
\item MR2 (Permutation of the attribute): If we permute the m attributes of all the samples and the test data, the result should remain unchanged. The appropriate value of K=3 (i.e.number of nearest neighbours) such that MR becomes a necessary property for IBK. The MR contains one source and follow-up test case.
\item MR3 (Addition of uninformative attributes): If we add some new feature that is equally associated with all classes, the predictions of the test data should not be changed. The appropriate value of K=3 (i.e.number of nearest neighbours) such that MR becomes a necessary property for IBK. The MR contains one source and follow-up test case.
\item MR4 (Consistence with re-prediction): Suppose we predict some test case t as class $l_{i}$ . If we append t to our training data and re-create the model, t should still be classified as class $l_{i}$. The appropriate value of K=3 (i.e.number of nearest neighbours) such that MR becomes a necessary property for IBK. The MR contains one source and follow-up test case.
\item MR5 (Additional training sample): For the source input, suppose we get the result $c_{t}$ = $l_{i}$ for the test case $t_{s}$. In the follow-up input, we duplicate all samples in S and L which have label $l_{i}$. The output of the follow-up test case should still be $l_{i}$. The appropriate value of K=3 (i.e.number of nearest neighbours) such that MR becomes a necessary property for IBK.  The MR contains one source and follow-up test case.
\item MR6 (Addition of classes by re-labeling sample): For some test cases not of class $l_{i}$, we switch the class label from x to x*. Then every test case predicted as class $l_{i}$ should still be predicted as class $l_{i}$ with the re-labeled samples. The appropriate value of K=3 (number of nearest neighbours) such that MR becomes a necessary property for kNN.  The MR contains one source and follow-up test case.
\item MR7 (Permutation of class labels): If we permute the order of the class labels with some random permutation p($l_{i}$) where $l_{i}$ is a class label, all test cases which were predicted as $l_{i}$ should now be predicted as p($l_{i}$). The appropriate value of K=1 (i.e.number of nearest neighbours) such that MR becomes a necessary property for IBK.  The MR contains one source and follow-up test case.
\item MR8 (Addition of informative attribute): If we add some new feature that is strongly associated with one class, $l_{i}$, then for every prediction that was class $l_{i}$, the prediction with this new attribute should also be class $l_{i}$. The appropriate value of K=1 (i.e.number of nearest neighbours) such that MR becomes a necessary property for IBK.  The MR contains one source and follow-up test case.
\item MR9 (Addition of classes by duplicating samples): Suppose we duplicate every class except for n, and give them all a new class. Then every test case predicted as class $l_{i}$ should still be predicted as class $l_{i}$ with the duplicated samples. The appropriate value of K=1 (i.e.number of nearest neighbours) such that MR becomes a necessary property for IBK.  The MR contains one source and follow-up test case.
\item MR10 (Removal of classes): If we remove some class $l_{i}$, the remaining predictions should remain unchanged. The appropriate value of K=1 (i.e.number of nearest neighbours) such that MR becomes a necessary property for IBK.  The MR contains one source and follow-up test case.
\item MR11 (Removal of samples): If we remove samples that have not been predicted as class $l_{i}$, then all cases which were predicted as $l_{i}$ should remain unchanged. The appropriate value of K=1 (i.e.number of nearest neighbours) such that MR becomes a necessary property for IBK.  The MR contains one source and follow-up test case.
\end{itemize}

\begin{figure}[h]
\centering
    \includegraphics[width=0.45\textwidth]{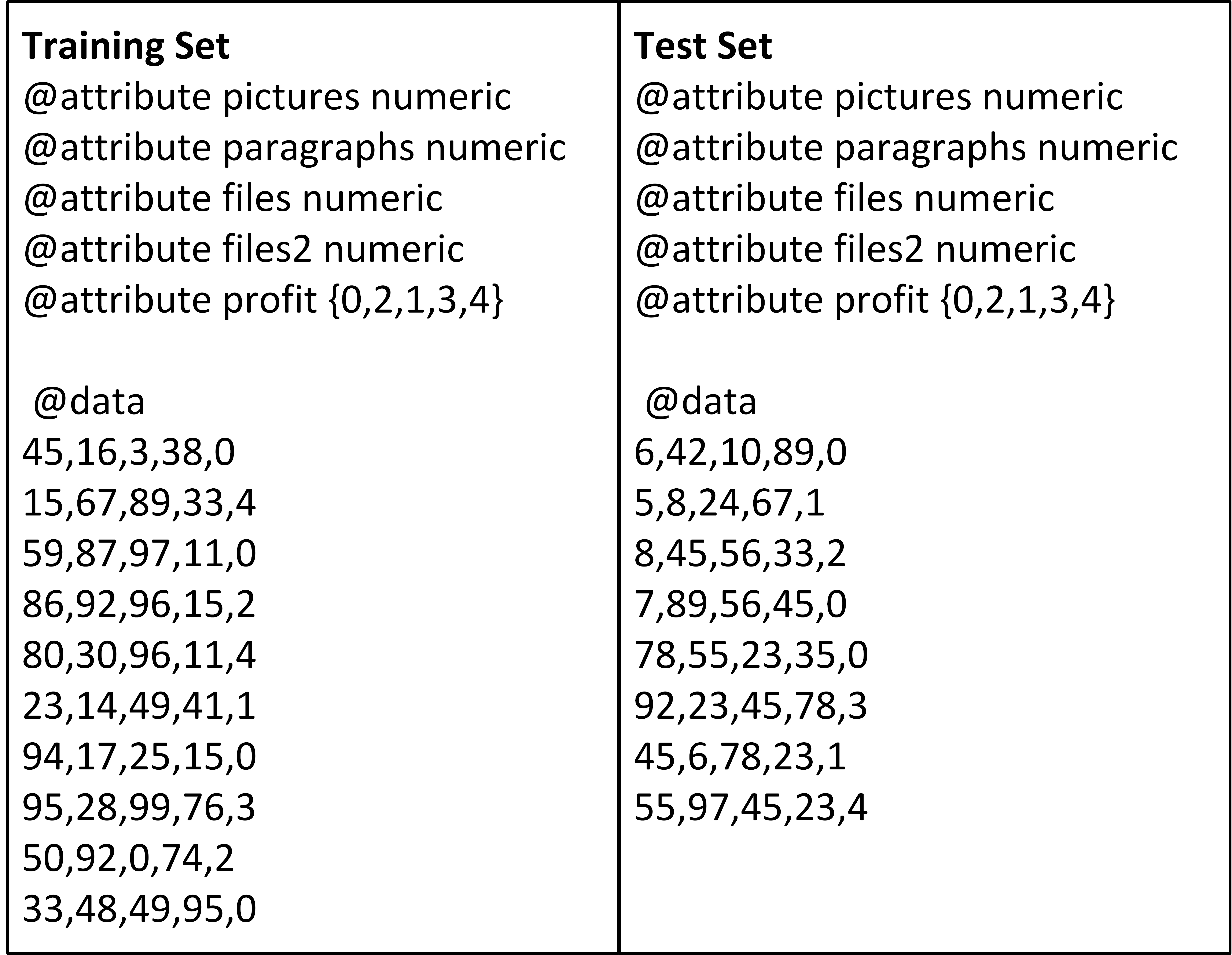}
    \caption{Sample Data Set for IBk Program}
    \label{fig:knnData}
\end{figure}

\subsection{MRs for LingPipe Program}
The MRs used for testing LingPipe program is discussed below and the MRs were modified from Srinivasan et al.~\cite{srinivasan2018quality}. In addition and removal MR for sentence and paragraph, sentence and paragraph boundary is maintained. Sentence boundaries are identified by scanning the text for sequences of characters separated by white space (tokens) containing one of the symbols !, . or ?~\cite{reynar1997maximum}.

\label{MR LingPipe}
\begin{itemize}
\item MR1 (Adding a sentence to another sentence): Given two sentences $S_{1}$ and $S_{2}$, we append $S_{2}$ to $S_{1}$ to form new sentence S$\prime$ such that the sentence boundary is maintained. We expect the union between the bio-entities of the sentence $S_{1}$ and sentence $S_{2}$ to be a subset of the bio-entities in the resultant sentence S$\prime$. The MR contains two source test cases (sentence 1 and sentence 2) and one follow-up test case(sentence)
\item MR2 (Adding a sentence to a paragraph): Given a sentence S and an index i, we add it to the position i of a paragraph P such it immediately follows sentence boundary. The resultant text of this addition is P$\prime$. We expect the union between the bio-entities of the sentence S and paragraph P to be a subset of the bio-entities in the resultant text P$\prime$. The MR contains two source test cases (sentence and paragraph) and one follow-up test case(paragraph)
\item MR3 (Adding a paragraph to an article): Given a paragraph P and an index i, we add this paragraph to the position i of an article A such it immediately follows paragraph boundary. The resultant text of this addition as A$\prime$. We expect the union between the bio-entities of the paragraph P and article A to be a subset of the bio-entities in the resultant text A$\prime$. The MR contains two source test cases (Paragraph and article) and one follow-up test case( Article).
\item MR4 (Adding a list of random words to another list): Given two list of random words $L_{1}$ and $L_{2}$, we append $L_{2}$ to $L_{1}$. The resultant list as L$\prime$. We expect the bio-entities of A$\prime$ to be the union of bio-entities extracted from $L_{1}$ and $L_{2}$. The MR contains two source test cases(word list1 and word list2) and follow-up test case (word list3).
\item MR5 (Removing a list of random words from a sentence): Given a sentence S, we remove a list of words L from S. The resultant text of this deletion as S$\prime$. We expect the bio-entities of S$\prime$ to be the subtraction of the bio-entities extracted of L from S. The MR contains two source test case (sentence and word) and one follow-up test case(sentence).
\item MR6 (Removing a sentence from a paragraph): Given a paragraph P and an index i, we remove a sentence S, starting from position i, from P such that the sentence boundary is maintained. We expect the difference between the bio-entities of the original paragraph and removed sentence to be a subset of the bio-entities in the resultant paragraph P$\prime$. The MR contains two source test case(Paragraph and sentence) and one follow-up test case(Paragraph).
\item MR7 (Removing a paragraph from an article): Given an article A, we remove a random paragraph P from A such that the paragraph boundary is maintained. The resultant text of this deletion as A$\prime$. We expect the difference between the bio-entities of the article A and removed paragraph to be a subset of the bio-entities in the resultant article A$\prime$. The MR contains two source test case(Article and Paragraph) and one follow-up test case(Article).
\item MR8 (Removing  some  words  from  a  list  of  random words): Given a list of random words $L_{1}$, we remove half of these words from the end of file. We refer to the removed part and the remaining part as $L_{2}$ and L$\prime$. We expect the bio-entities of L$\prime$ to be the subtraction of the bio-entities of $L_{1}$ and $L_{2}$. The MR contains two source test case (list of thousand random words $L_{1}$ and list of five hundred words $L_{2}$) and one follow-up test case( list of five hundred word $L_{2}$). 
\item MR9 (Shuffling paragraphs of an article): Given an article A, we shuffle all the paragraphs of A. The new resultant text as $A\textprime$. We expect the bio-entities of the article A to be subset of the resultant text $a\textprime$, although the position of the bio-entities can vary. The MR contains one source test case (Article) and one follow-up test case (Shuffled Article).
\item MR10 (Shuffling a list of random words): Given a list of random words L, we shuffle all words and create a new list of words L. We create a new list of words $L\textprime$. We expect the bio-entities of $a\textprime$ to be equal to the bio-entities extracted from L. The MR contains one source test case( random words) and one follow-up test case( shuffled words). 
\end{itemize}